\documentclass{article}

\usepackage{arxiv}

\usepackage[utf8]{inputenc} 
\usepackage[T1]{fontenc}    
\usepackage{hyperref}       
\usepackage{url}            
\usepackage{booktabs}       
\usepackage{amsfonts}       
\usepackage{nicefrac}       
\usepackage{microtype}      
\usepackage{lipsum}		
\usepackage{graphicx}
\usepackage{natbib}
\usepackage{doi}

\title{EmoDiarize: Speaker Diarization and Emotion Identification from Speech Signals using Convolutional Neural Networks}

\author{Hanan Hamza\\
	School of Digital Sciences\\
	Kerala University of Digital Sciences-\\Innovation and Technology\\
	Thiruvananthapuram, India \\
	\texttt{hanan.ds22@duk.ac.in} \\
        \And
        Fiza Gafoor M\\
	School of Digital Sciences\\
	Kerala University of Digital Sciences-\\Innovation and Technology\\
	Thiruvananthapuram, India \\
	\texttt{fiza.ds22@duk.ac.in} \\
	\And
        Fathima Sithara E\\
	School of Digital Sciences\\
	Kerala University of Digital Sciences-\\Innovation and Technology\\
	Thiruvananthapuram, India \\
	\texttt{fathima.ds22@duk.ac.in} \\
        \And
        Gayathri Anil\\
	School of Digital Sciences\\
	Kerala University of Digital Sciences-\\Innovation and Technology\\
	Thiruvananthapuram, India \\
	\texttt{gayathri.ds22@duk.ac.in} \\
	\And
	V. S. Anoop \\
        Applied NLP Research Lab\\
	School of Digital Sciences\\
	Kerala University of Digital Sciences-\\Innovation and Technology\\
	Thiruvananthapuram, India \\
	\texttt{anoop.vs@duk.ac.in} \\
}

\date{}

\renewcommand{\shorttitle}{EmoDiarize: Speaker Diarization and Emotion Identification from Speech}

\hypersetup{
}

\begin{document}
\maketitle

\begin{abstract}
 In the era of advanced artificial intelligence and human-computer interaction, identifying emotions in spoken language is paramount. This research explores the integration of deep learning techniques in speech emotion recognition, offering a comprehensive solution to the challenges associated with speaker diarization and emotion identification. It introduces a framework that combines a pre-existing speaker diarization pipeline and an emotion identification model built on a Convolutional Neural Network (CNN) to achieve higher precision. The proposed model was trained on data from five speech emotion datasets, namely, RAVDESS, CREMA-D, SAVEE, TESS, and Movie Clips, out of which the latter is a speech emotion dataset created specifically for this research.  The features extracted from each sample include Mel Frequency Cepstral Coefficients (MFCC), Zero Crossing Rate (ZCR), Root Mean Square (RMS), and various data augmentation algorithms like pitch, noise, stretch, and shift. This feature extraction approach aims to enhance prediction accuracy while reducing computational complexity. The proposed model yields an unweighted accuracy of 63\%, demonstrating remarkable efficiency in accurately identifying emotional states within speech signals.
\end{abstract}


\section{Introduction}
Speech Emotion Recognition (SER) is a specialized branch of Natural Language Processing (NLP) that centers around interpreting emotions conveyed through spoken language \cite{al2023speech}. The human voice carries a wealth of emotional information in its pitch, intensity, rhythm, and timbre, which makes it a valuable resource for understanding the emotional state of an individual. Emotion recognition from audio samples is a valuable application of machine learning and signal processing, with numerous potential use cases in areas like entertainment, market research, mental health assessment, and more \cite{wagner2023dawn}. This process involves analyzing audio data, often in the form of speech or vocal expressions, to determine the emotional state of the speaker.  It uses machine learning and deep learning techniques to discern the emotional state of a person from their voice, whether it is in spoken language, song, or other vocalizations \cite{de2023ongoing}. 

The practical applications of this technology are far-reaching. In entertainment, emotion recognition can transform how we interact with media, making it more immersive and responsive \cite{daneshfar2023octonion}. In marketing and market research, it can provide invaluable insights into consumer sentiment and preferences, enhancing product development and advertising strategies \cite{kaur2023trends}. Additionally, the field of mental health holds the potential to assist in early diagnosis, monitoring, and treatment, offering a new dimension to patient care \cite{anoop2023public}\cite{varghese2022deep}. However, amidst these promises, we must navigate challenges related to accuracy, privacy, and the ethical implications of emotion recognition technology\cite{anoop2023sentiment}. This report aims to navigate this multifaceted landscape, shedding light on the methods, opportunities, and responsibilities inherent in the fascinating world of emotion recognition from audio.

Speaker diarization is the process of segmenting audio recordings by speaker labels, and it effectively tells us "Who spoke when?” in the audio recording \cite{serafini2023experimental}. It faces challenges, particularly in situations with multiple speakers talking simultaneously, varying speech patterns, background noise, or low-quality audio recordings \cite{wu2023semi}. It has a wide range of applications, including transcribing multi-speaker conversations accurately, indexing large audio and video archives, and enhancing the performance of speech recognition systems when multiple speakers are involved \cite{maiti2023eend}. Various methods have been employed for speaker diarization over the years. Traditional techniques include Gaussian Mixture Models (GMMs) and Hidden Markov Models (HMMs). More recently, deep learning-based methods, such as recurrent neural networks (RNNs) and convolutional neural networks (CNNs), have shown promise in improving diarization accuracy, especially in complex scenarios \cite{park2022review}\cite{sreelakshmideep}.
 
This paper presents a machine-learning model that performs speaker diarization and emotion identification from speech signals using CNN. The structure of the paper is as follows: Section \ref{ref_related} reviews related research, while Section \ref{ref_materials} defines important concepts like CNN and speaker diarization and introduces the relevant datasets. Section \ref{ref_propapp} elaborates on the proposed approach, followed by a detailed account of the experiments conducted in Section \ref{ref_experiments}. Section \ref{ref_results} then presents the results obtained. Finally, Section \ref{ref_conclusion} presents the conclusion and discusses potential directions for future research.

\section{Related studies}
\label{ref_related}
The significance of Speech Emotion Recognition lies in its capacity to enhance human-computer interactions, provide fresh insights into human behavior, and find applications in various fields, ultimately improving the quality of services and experiences. This importance of SER has piqued significant interest, leading to numerous studies and surveys. This section reviews recent publications and research closely aligned with this study area. \cite{khalil2019speech} discussed traditional SER methods and their limitations, highlighting the need for more advanced approaches, particularly those involving deep learning. The paper thoroughly examines various deep learning techniques like Deep Boltzmann Machines (DBM), Recurrent Neural Networks (RNN), Convolutional Neural Networks (CNN), and Auto Encoders (AE) and for each of these techniques, provides a detailed analysis of their advantages and disadvantages within the context of SER. This review underscores the growing importance of deep learning in SER research while acknowledging its limitations. A similar study was carried out \cite{abbaschian2021deep}, comparing the existing approaches and databases used in SER. They focused on two main areas: deep learning methods and conventional machine learning techniques, conducting a comprehensive comparison of various practical neural network approaches to speech emotion recognition. \cite{lieskovska2021review} discusses the recent incorporation of attention mechanism into DNN architecture to emphasize emotionally significant information. It reviews recent developments in SER and investigates how the different attention mechanisms affect SER performance. The study concludes with a comprehensive comparison of system accuracies using the IEMOCAP benchmark database. 

Choosing the right feature extraction techniques in SER is crucial as it determines how effectively relevant information is captured from speech signals. \cite{gupta2020emotion} investigates various such techniques, focusing on prosodic and spectral features like Mel Frequency Cepstral Coefficients (MFCC), Linear Prediction Cepstral Coefficients (LPCC), and Linear Prediction Coefficients (LPC). The study concludes that MFCC is the most effective for recognizing emotional content in speech. Existing classification techniques are applied to identify human emotions, with a detailed comparative analysis based on statistical and mathematical results. The paper also introduces an optimal model, Depthwise Separable Convolutional Neural Network (DSCNN), designed for raw spectrogram data. \cite{issa2020speech} introduces a novel architecture that directly processes sound data, extracting various acoustic features from audio files, and employs a one-dimensional Convolutional Neural Network (CNN) for emotion identification on publicly available datasets RAVDESS, EMO-DB and IEMOCAP. Using an incremental model refinement, their best-performing model achieves state-of-the-art results in RAVDESS and IEMOCAP, and performs competitively in the EMO-DB dataset. \cite{anvarjon2020deep} introduces an SER model that utilizes a CNN approach to learn deep frequency features using a specific rectangular filter and modified pooling strategy. The model is trained on frequency features extracted from speech data and tested for emotion prediction. Evaluation on two benchmarks, IEMOCAP and EMO-DB speech datasets, results in high recognition accuracies, demonstrating that the proposed system outperforms state-of-the-art SER systems.

To address the challenge of inadequate datasets for training deep learning models in SER, \cite{abdelhamid2022robust} proposed a data augmentation algorithm to enrich the dataset by introducing carefully designed noise fractions. In addition, they introduce an optimized deep learning model that combines a Convolutional Neural Network (CNN) and a Long Short-Term Memory (LSTM) layer to capture both local and long-term correlations in speech samples along with hyperparameters fine-tuned for improved recognition results. The learning rate and label smoothing regularization factor are optimized using a stochastic fractal search-guided whale optimization algorithm. Experimental results on speech emotion datasets IEMOCAP, EMO-DB, RAVDESS, and SAVEE demonstrate the superior performance of the proposed approach, achieving high recognition accuracies. The recognition accuracy of speech emotion recognition systems can also be improved using alternate methods. This is demonstrated in \cite{koduru2020feature} using different feature extraction algorithms. It emphasizes preprocessing audio samples where noise is removed using filters and employs Mel Frequency Cepstral Coefficients (MFCC), Discrete Wavelet Transform (DWT), pitch, energy, and Zero Crossing Rate (ZCR) algorithms for feature extraction. Global feature selection reduces redundant information, and machine learning classification algorithms are applied to identify emotions. The study validates these algorithms for universal emotions, namely Anger, Happiness, Sadness, and Neutral. 

\section{Materials and Methods}
\label{ref_materials}
\subsection{Convolutional Neural Networks (CNN)}
\label{ref_CNNmodel}
\begin{figure}
      \centering
      \includegraphics[width=0.5\linewidth]{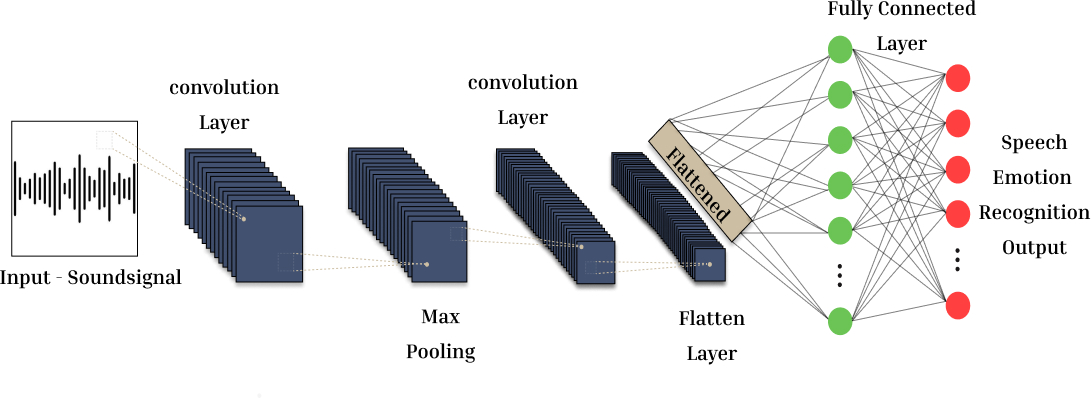}
      \caption{CNN network architecture}
      \label{fig:cnn_architecture}
  \end{figure}

Convolutional Neural Networks (CNNs), often called ConvNets, constitute a prominent category within the domain of deep learning methodologies characterized by their feed-forward architecture. These neural networks have emerged as potent tools, particularly well-suited for the processing of structured grid data. While initially devised for computer vision applications, their utility extends beyond visual tasks to encompass domains like speech analysis and natural language processing. CNNs excel in scenarios involving structured grid data, encompassing many data types, including images and spectrograms. This ability comes from the fact that CNNs are designed to work like our visual system. They are inspired by how our eyes and brains hierarchically see things and process information. The overall architecture of the CNN model is shown in Figure \ref{fig:cnn_architecture}, which illustrates the layered architecture of a fundamental CNN network, providing a comprehensive overview of the network's structural components and their interconnections. The architectural composition of a Convolutional Neural Network (CNN) is foundational to its functionality, comprising several key components:
\begin{itemize}
    \item \textbf{Convolutional Layer}: Convolutional layers are the building blocks of CNNs. Convolutional operations are used to extract local features out of the input data. Multiple filters make up each layer, which move across the input data to capture spatial hierarchies of features. The convolution operation can be described mathematically as: 
    \begin{equation}
        Z[i,j]=(X*W)[i,j]+b
    \end{equation}
 
    In this equation, 'X' represents the input data, typically a spectrogram, 'W' signifies the weight matrix of the convolutional filter, and 'b' accounts for the bias term. The outcome 'Z[i,j]' denotes the weighted summation of input values at the position (i,j) and is obtained by convolving the filter 'W' with input data 'X' and factoring in the bias 'b.'

    \item \textbf{Pooling Layers}: Pooling layers, which are frequently interspersed with convolutional layers, are vital for lowering spatial dimensions while preserving important data. Common pooling methods include max-pooling and average-pooling, which downsample feature maps by choosing the maximum or average value within certain geographic areas.

    \item \textbf{Fully Connected Layers}: Fully Connected layers at the CNN architecture's core allow for high-level decision-making and global context aggregation. These layers establish connections between every neuron and every neuron in the layer above, enabling the model to understand intricate correlations between characteristics that were retrieved by earlier levels.
\end{itemize}
\subsection{Speaker Diarization}
\label{ref_SpeakerDiarization}
Speaker diarization is a vital process in the realm of audio analysis, aimed at segmenting an audio stream containing human speech into coherent portions based on the identity of each speaker. This technique not only enhances the comprehensibility of automated speech transcriptions but also facilitates the organization of the audio stream into distinct speaker segments. Additionally, when integrated with speaker recognition systems, diarization identifies speakers, addressing the fundamental question of "who spoke when." In this context, PyAnnote Audio (available at \url{https://github.com/pyannote/pyannote-audio}) emerges as a noteworthy open-source toolkit implemented in Python, meticulously crafted for speaker diarization. Version 2.1 marks a significant evolution, introducing substantial improvements to the default speaker diarization pipeline, which comprises three pivotal stages. Firstly, there's the speaker segmentation phase, which operates on a short sliding window. Next, the toolkit incorporates neural speaker embedding, allowing for the extraction of unique speaker characteristics. Lastly, the process culminates with agglomerative clustering, which effectively groups speaker embeddings, facilitating the identification and separation of individual speakers. This latest version, PyAnnote Audio Speaker Diarization (Version 2.1), represents a substantial advancement in diarization and offers a more refined and effective means of tackling the challenge of "who spoke when." Its enhancements have the potential to greatly benefit various applications, including transcription systems and speaker identification tasks. \cite{bredin2023pyannote}

\subsection{Datasets}
\label{ref_Datasets}
The following datasets were used to build the machine-learning model. The first four are publicly available datasets. The final dataset was created specifically for the purpose of this project.
\begin{itemize}
    \item \textbf{RAVDESS (The Ryerson Audio-Visual Database of Emotional Speech and Song)}: This dataset contains 7,356 files (total size: 24.8 GB). The database contains 24 professional actors (12 female, 12 male), vocalizing two lexically matched statements in a neutral North American accent. RAVDESS is incredibly rich in sample variants; also, each emotion is rendered in two different intensities and with both a normal and singing voice. Speech includes calm, happy, sad, angry, fearful, surprise, and disgust expressions, and song contains calm, happy, sad, angry, and fearful emotions. This is one of RAVDESS's most important qualities; only a few data sets can claim to contain it. To build this specific model, only the 1,440 WAV files containing speech audio were used. \cite{issa2020speech}
    \item  \textbf{CREMA-D (Crowd-Sourced Emotional Multimodal Actors Dataset)}: This data set is made for multimodal emotion recognition tasks. The dataset consists of 7,442 original clips from 91 actors from various ethnic backgrounds. These clips are from 48 male and 43 female actors between the ages of 20 and 74 coming from various races and ethnicities (African American, Asian, Caucasian, Hispanic, and Unspecified). Actors spoke from a selection of 12 sentences. The labels were created using the crowdsourcing of 2,443 raters. Each sample in this dataset contains two ratings, one for the emotion category and the other for intensity. The sentences were presented using one of six different emotions (Anger, Disgust, Fear, Happy, Neutral, and Sad) and four different emotion levels (Low, Medium, High, and Unspecified). \cite{cao2014crema}
    \item \textbf{TESS (Toronto Emotional Speech Set)}: This is an acted dataset created primarily to investigate the influence of aging on the capacity to recognize emotions. There is a set of 200 neutral target sentences spoken in the carrier phrase "Say the word" by two actresses (aged 26 and 64 years), and recordings were made of the set portraying each of seven emotions (anger, disgust, fear, happiness, pleasant surprise, sadness, and neutral). 56 undergraduate students were asked to identify emotions from phrases to categorize the dataset. Following the identification task, statements with greater than 66\% confidence were chosen to be included in the dataset. There are 2,800 data points (audio files) in total. All the audio files are in WAV format. \cite{dupuis2010toronto}
    \item  \textbf{SAVEE (Survey Audio–Visual Expressed Emotion)}: This dataset was collected from four native English male speakers (designated as DC, JE, JK, and KL), postgraduate students and researchers at the University of Surrey ranging in age from 27 to 31 years. Anger, contempt, fear, pleasure, sorrow, and surprise have all been classified as distinct emotions in psychology. A neutral category is also included, resulting in records of 7 emotion types. The text material consisted of 15 TIMIT sentences per emotion: 3 common, 2 emotion-specific, and 10 generic lines that were phonetically balanced and distinct for each emotion. To make 30 neutral sentences, the 3 common and 2*6 = 12 emotion-specific phrases were registered as neutral. This resulted in 120 utterances per speaker \cite{ryumina2022search}.
    \item  \textbf{Movie clips}: The dataset was compiled by randomly selecting audio clips from movies and then annotating them based on the expressed emotions. It has a total of 166 clips. The emotions identified are anger, happy, fear, disgust, sad, surprise, and neutral. 
\end{itemize}

\section{Proposed Approach}
\label{ref_propapp}
\begin{figure}[h!]
    \centering
    \includegraphics[width=0.5\linewidth]{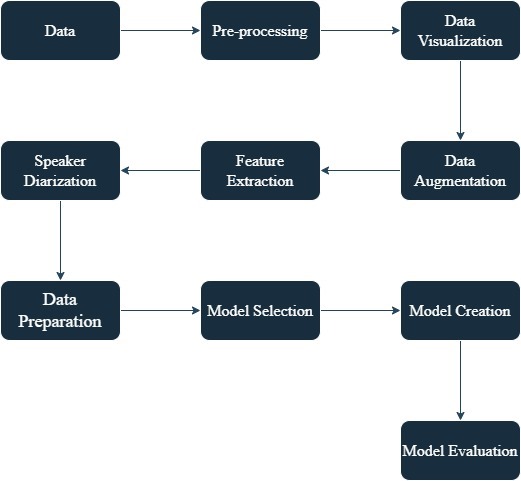}
    \caption{The overall workflow of the proposed speaker diarization and speech emotion recognition approach} 
    \label{fig:flowchart}   
\end{figure}

\subsection{Dataset collection and pre-processing}
\label{ref_data}
\begin{figure}[h!]
    \centering
    \includegraphics[width=0.5\linewidth]{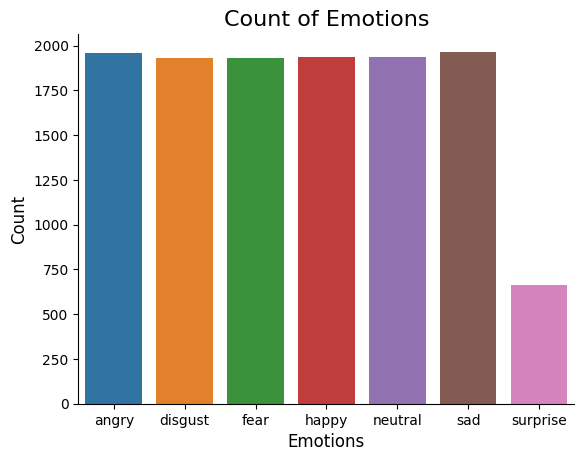}
    \caption{Count of emotion}
    \label{fig:emotionsgraph}
\end{figure}

Out of the five datasets used, four are publicly available: RAVDESS, CREMA-D, SAVEE, and TESS. The final one, 'Movie clips,' was created by taking audio clips from different movies specifically for the purpose of this project. The specifications have been described in Section \ref{ref_Datasets}. All the datasets are integrated into a single dataset where each audio file is classified into one of seven emotions - sad, fear, disgust, neutral, happy, surprise, and anger. A graph of the seven emotions is plotted against the count of emotions as shown in Figure \ref{fig:emotionsgraph}.
\subsection{Data Augmentation}
\label{ref_DataAugmentation}
Data augmentation is a technique used in deep learning to improve the quality of the data being used to train a model. It does so by adding small variations to existing data samples. It helps prevent the models from overfitting and improves the model accuracy \cite{ferreira2022survey}. The techniques used here are as follows:
\begin{itemize}    
\item Noise: Add random noises to the audio recordings
\item Shifting: Shift the audio left or right with random seconds.
\item Stretching: Stretch the audio while preserving the audio contents.
\item Changing pitch: Randomly change the pitch of the audio
\end{itemize}
\subsection{Feature Extraction}
\label{ref_FeatureExtraction}
Feature extraction is a crucial step in analyzing and establishing relationships between various elements. Audio data, in its raw form, cannot be directly fed into the models. So it is necessary to convert it into a format that models can understand, and this is where feature extraction comes into play. The audio signal is a 3-dimensional signal in which the three axes represent time, amplitude, and frequency. By using the sample rate and sample data, several transformations can be performed to extract valuable features out of it. 
\begin{itemize}
\item Zero Crossing Rate: The rate of sign-changes of the signal during the duration of a particular
Frame.
\item MFCC: Mel Frequency Cepstral Coefficients form a cepstral representation where the
frequency bands are not linear but distributed according to the mel scale.
\item RMS (Root Mean Square) value: It is the square of the arithmetic mean or the square of the
function. \cite{aggarwal2022two}
\end{itemize}
\subsection{Speaker Diarization}
Speaker diarization is the process of segmenting the human-speaking audio stream into separate parts based on the identity of each speaker. By dividing the audio into speaker turns, this method makes computerized speech transcriptions easier to understand. As discussed in Section \ref{ref_SpeakerDiarization} \cite{park2022review}
\subsection{Data Preparation}
\label{ref_dataprep}
Splitting the dataset is a common practice in machine learning to assess the performance of the models on unseen data and avoid overfitting. Hence, it enables the development and evaluation of machine learning models. Proper data splitting helps assess a model’s generalization performance and ability to make accurate predictions on new and unseen data. It takes several parameters to control how the data should be split. ‘X’ typically contains the independent variables or features that the model uses to make predictions. ‘Y’ typically contains the corresponding target values or labels that your model is trying to predict. The random state parameter is important for reproducibility. It sets a seed for the random number generator, ensuring that the same random split will be generated every time you run this code with the same seed. In this case, the random state is 42. The test size parameter determines the proportion of the data allocated to the testing set. In this case, 20\% of the data will be used for testing, and the remaining 80\% will be used for training. Shuffling is usually set to ‘True’ to prevent any inherent order or bias in the data from affecting the model's performance. 
\subsection{Model Selection}
\label{ref_modelselection}
The emotion expressed in each audio segment obtained after speaker diarization is identified in
this step. We employed multiple models, namely LSTM, CNN, and CLSTM, for emotion prediction.
After testing these models, it was determined that the CNN model provided the most accurate
predictions, and as a result, we selected the same. In this case, a sequential model is employed to construct the CNN.
\subsection{Model Creation}
\subsubsection{CNN Model}
\label{ref_cnn}
Convolutional Neural Networks (CNNs) are a fundamental component of machine learning. These networks are interconnected layers designed to learn and extract intricate patterns and features from images automatically. Convolutional layers apply learnable filters to scan the input image, while activation layers introduce non-linearity, enabling the network to capture complex relationships. Pooling layers reduce the spatial dimensions, and fully connected layers make the final classification.By training on labeled data and adjusting internal parameters, CNNs excel at recognizing objects, shapes, and patterns in images, making them a cornerstone technology in computer vision and image analysis within machine learning. \cite{bhatt2021cnn}
\begin{itemize}
\item \textbf{Sequential Model}
A sequential model in machine learning provides a structured framework for organizing and applying processing steps in a specific sequence, ensuring that data flows through the model in a predetermined order. The term "sequential" refers to the specific way in which layers are stacked in the model, where one layer follows another in a sequential order. Sequential models are commonly used in various neural networks, including CNN. CNNs often have a sequence of layers that perform operations like convolution, pooling, flattening, and fully connected layers. These layers are typically arranged sequentially in the order they are applied to the input data, as explained in Section \ref{ref_CNNmodel}
\end{itemize}

\section{Experiments}
\label{ref_experiments}
This project was done in a system with an Intel core i5 processor having Windows 11. We have executed the code in Visual Studio code with Python environment 3.9. First, we have collected 5 datasets, of which 4 are available on the internet, and the 5th dataset was created especially for this project. Then we integrated the 5 datasets into a single dataset. The next step is data augmentation.
\begin{itemize}
    \item \textbf{Data Augmentation:} The process in which our training dataset is converted to a new synthetic data sample by adding small perturbations is called Data Augmentation. It is done by applying noise injection, shifting time, changing pitch. It is to make the model invariant to those perturbations and enhance its generalization ability. (Refer Section \ref{ref_DataAugmentation}.)
    \item \textbf{Feature Extraction:} During the features extraction phase, three features – Zero Crossing Rate (ZCR),Mel Frequency Cepstral Coefficient (MFCC), and Root Mean Square Value (RMS) – are extracted. The duration for the feature extraction process is set to 2.5 seconds and the offset to 0.6 seconds to accommodate the varying data lengths. A total of 22 features are extracted from each audio, containing one ZCR value, one RMS value, and 20 MFCC coefficients (Refer Section \ref{ref_FeatureExtraction})
     \item \textbf{Speaker Diarization:} Speaker diarization is the process of identifying and separating individual speakers in an audio stream so that, in the automatic speech recognition (ASR) transcript, each speaker's utterances are identified and separated. Speaker diarization is used in scenarios where multiple speakers are involved. Here, a pre-trained speaker diarization pipeline is used to identify and separate two speakers from a given set of audio samples. The integrated dataset is run through the pre-trained model, and the output contains utterances of each speaker separately, along with their start time and end time. The result of the diarization process is represented in the RTTM (Rich Transcription Time Marked) format.
    \item \textbf{Splitting the dataset for machine learning:} This is a common practice in machine learning to assess the performance of the models on unseen data and avoid overfitting. Hence, it enables the development and evaluation of machine learning models. Proper data splitting helps assess a model’s generalization performance and its ability to make accurate predictions on new and unseen data. It takes several parameters to control how the data should be split. ‘X’ typically contains the independent variables or features that the model uses to make predictions. ‘Y’ typically contains the corresponding target values or labels that your model is trying to predict. The random state parameter is important for reproducibility. It sets a seed for the random number generator, ensuring that the same random split will be generated every time you run this code with the same seed. In this case, the random state is 42. The test size parameter determines the proportion of the data allocated to the testing set. In this case, 20\% of the data will be used for testing, and the remaining 80\% will be used for training. Shuffling is usually set to ‘True’ to prevent any inherent order or bias in the data from affecting the model's performance.
     \item \textbf{CNN Model Architecture:} We have selected CNN architecture for our model building. We have used a sequential model to implement CNN architecture. A sequential model is a model in which layers are added one after the other. It starts with a convolutional layer with 512 filters, a kernel size of 5, and 'same' padding. The activation function used here is ReLU- Rectified Linear unit. It works as follows: 
        \begin{itemize}
            \item If input to the function is positive, it outputs that positive value. 
            \item If the input value is negative, then it outputs zero.
        \end{itemize}
 A batch normalization is applied after this convolutional layer.  it works by normalizing the inputs to a layer in a mini-batch of data, adjusting them to have zero mean and unit variance. Then, there is a max-pooling layer with a pool size of 5 and a stride of 2. This reduces the spatial dimensions of the feature maps. Another set of convolutional layers, batch normalization layers, and max-pooling layers is added. These layers help the model extract and learn hierarchical features from the input data. A dropout layer with a dropout rate 0.2 is added after the second max-pooling layer. Dropout is a regularization technique to prevent overfitting. Again 2 sets of convolutional, batch normalization, and max-pooling layers are added. Also, a dropout layer with a dropout rate of 0.2 is added. this is repeated again. After several convolutional and max-pooling layers, there's a flattened layer that transforms the 2D feature maps into a 1D vector. Two dense (fully connected) layers are added. The first dense layer has 512 units and ReLU activation. Batch normalization is applied after the first dense layer.
The second dense layer has 7 units (assuming it's a classification task with 7 classes) and uses the softmax activation function, which is typical for multi-class classification. Then training process is configured. It uses the Adam optimizer, categorical cross-entropy loss (common for multi-class classification), and accuracy as the evaluation metric. Then the model is evaluated.
\end{itemize}
\section{Results and Discussions}
\label{ref_results}
This section discusses the results obtained from the experiment conducted using our proposed framework which is followed by a detailed discussion.
\begin{figure}[h!]
    \centering
    \includegraphics[width=0.5\linewidth]{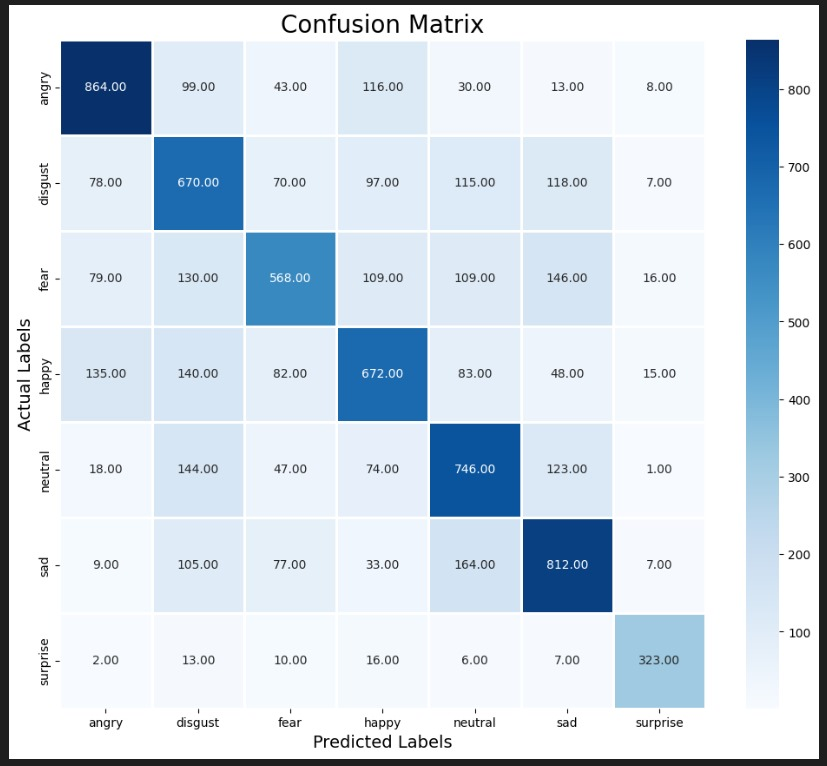}
    \caption{Confusion matrix} 
    \label{fig:confusionmatrix}
\end{figure}
The model is better at predicting surprise and irrational emotions. It makes sense because the audio files representing these emotions are unique from all other audio files in terms of pitch, speed, and other factors. On our test data, we managed to attain an overall accuracy of 61\%. By employing additional augmentation strategies and different feature extraction approaches, we can further enhance it. Correct predictions are shown in the top-left to bottom-right diagonal cells of the confusion matrix in Figure \ref{fig:confusionmatrix}. The better the model performs, the darker these cells are. The diagonal cells are darker than the others, indicating that the model is functioning properly and making accurate predictions. \cite{yang2019human}

\begin{figure}
    \centering
    \includegraphics[width=0.5\linewidth]{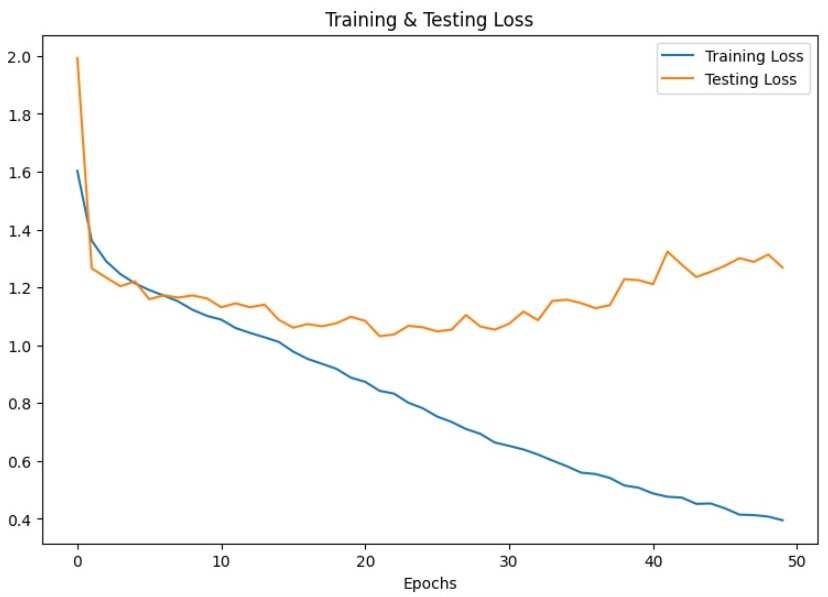}
    \caption{Graph showing training vs testing loss}
    \label{fig:lossplot}
\end{figure}

\begin{figure}
    \centering
    \includegraphics[width=0.5\linewidth]{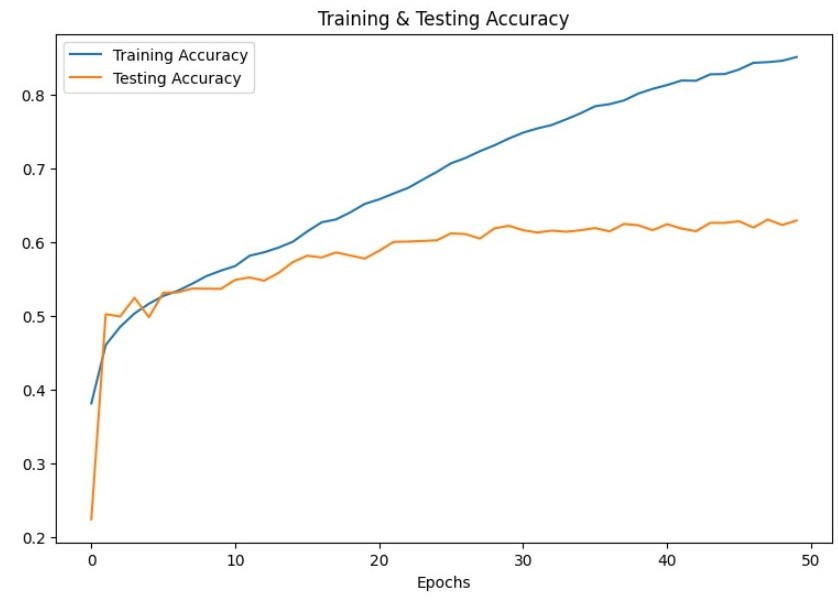}
    \caption{Graph showing training vs testing accuracy}
    \label{fig:accuracyplot}
\end{figure}

The classification report obtained is given in Table \ref{ref_classificationreport}:

\begin{table}[h!]
\centering
\caption{The precision, recall, f1-score produced by the proposed approach for different classes}
\begin{tabular}{|l|l|l|l|l|}
\hline
\textit{} &Precision  &Recall &F1-score  &Support  \\ \hline
         Angry & 0.73 &0.74  &0.73  & 1173 \\ \hline
          Disgust&0.51  &  0.58&0.55  &1155  \\ \hline
          Fear&0.63  &0.49  &0.55  &1157  \\ \hline
          Happy& 0.60 &0.57  &0.59  &1175  \\ \hline
          Neutral&  0.60& 0.65 &  0.62&1153  \\ \hline
          Sad&  0.64&  0.67& 0.66 & 1207 \\ \hline
          Surprise& 0.86 &0.86  &0.86  &377  \\ \hline
          \textbf{accuracy}&  &  &0.63  &  7397\\ \hline
          \textbf{macro avg}&0.65  & 0.65 & 0.65 & 7397 \\ \hline
          \textbf{weighted avg}& 0.63 & 0.63 & 0.63 & 7397 \\ \hline
\end{tabular}
\label{ref_classificationreport}
\end{table}
\begin{figure}
    \centering
    \includegraphics[width=0.5\linewidth]{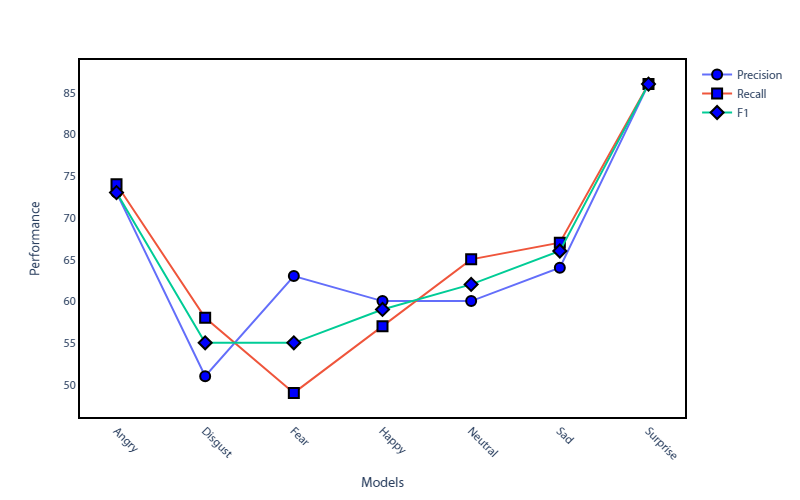}
    \caption{Precision, recall, f1-score of the proposed approach on different classes}
    \label{fig:linegraph}
\end{figure}
The model accurately predicts the emotion "surprise" with a precision of 0.86, the highest precision value in the classification report. This indicates that most "surprise" feelings are correctly predicted. Additionally, the emotion's F1-score and recall value are also 0.86. The accuracy of the model is calculated by:
\begin{equation}
    Accuracy = (TP + TN) / (TP + TN + FN + FP)
\end{equation}
where TP is True Positive, TN is True Negative, FN is False Negative, and FP is False Positive. Here, the accuracy obtained is 63\%. The training vs testing loss is plotted in the graph shown in Figure \ref{fig:lossplot}
Training loss is initially fairly substantial at epoch 1. This is due to the model's inability to recognize any patterns in the data yet. The training loss gradually diminishes as training goes on (from epochs 2 to 50). This shows that the model is growing more adept at identifying patterns in the training data and at fitting the data more accurately. Similarly, because the model has not yet been exposed to the testing data, the testing loss likewise begins at higher values during epoch 1. However, the testing loss initially diminishes as training continues (from epochs 2 to 50). This is a good indicator because it shows that the model is successfully generalizing to new data. Although there may be minor swings in the testing loss graph, there is a little rising tendency. As a result, the model may become excessively specialized in fitting the training data and may not perform as well on fresh data, raising the possibility of overfitting. The training vs testing accuracy graph is shown in Figure \ref{fig:accuracyplot}.
Since the model has not yet acquired the characteristics of the data, the training accuracy is generally low at the beginning of the training (epoch 1). The training accuracy progressively rises from epoch 2 to 50 as the training proceeds. It shows that the model is more accurate in foreseeing the feelings gleaned from the practice dataset. The testing accuracy is comparable to the training accuracy low when the training first began. The testing accuracy improves as the instruction goes on. Here is a favorable indication that the model is mastering feature recognition. It is evident that at the end of the training, the testing accuracy plot virtually reaches a stable state. This indicates that the model has learned the underlying patterns in the data and is generalizing to previously unexplored data. 
\section{Conclusion and Future Work}
\label{ref_conclusion}
Emotions are essential to communication because they enable us to express complicated feelings verbally. The goal of our study was to develop a smart system that can recognize these emotions in speech. We created this system by fusing two technologies: one that can distinguish between speakers in audio recordings and another that can foretell each speaker's emotions. 
This system has several applications. For instance, it can assist businesses in understanding clients' emotions during encounters, allowing them to enhance services in real time. It offers a deeper understanding of customer emotions and goes beyond straightforward satisfaction surveys. Therapists can utilize this approach to more effectively comprehend their patients' feelings during therapy sessions in psychological care, enabling more focused and sympathetic interventions. This could improve the standard of mental health care while also potentially accelerating the healing process for patients. Our experiment demonstrates the intriguing potential of integrating audio separation and emotion prediction to enhance human-technology interaction. \cite{huang2014speech}. Exciting possibilities lie in the future of speaker diarization, which separates and identifies speakers in audio recordings. With modern technology like deep learning, improved accuracy is also to be expected, making it more dependable in varied scenarios. Teleconferencing and real-time applications like live transcription will function more effectively. Speaker diarization will become more advanced to support several languages, incorporate various media kinds (text, video), and have applications in healthcare, education, and entertainment. It will be essential to address privacy issues, and adjusting to complicated audio situations with multiple speakers is a top priority. The future of this technology will also be shaped by customization for particular circumstances, standard evaluation measures, and integration with language and emotion recognition, making it more accessible and adaptable for various applications.
\bibliographystyle{unsrtnat}
\bibliography{references} 
\end{document}